\documentclass[twocolumn,preprintnumbers,amsmath,amssymb,superscriptaddress,nofootinbib,longbibliography, prb]{revtex4-2}

\usepackage{graphicx}
\usepackage{bm}
\usepackage{dsfont}
\usepackage[usenames,dvipsnames]{xcolor}
\usepackage{pstricks}
\usepackage[tight]{subfigure}
\usepackage{verbatim}
\usepackage{units}
\usepackage{multirow}
\usepackage{mathrsfs}
\usepackage{leftidx}
\usepackage{xspace}
\usepackage{diffcoeff}  
\usepackage{bbm}
\usepackage{amsfonts,amssymb,amsmath}
\usepackage{mathtools}

\usepackage{array} 
\usepackage{tabularx}
\usepackage[normalem]{ulem}
\usepackage{bbold}
\usepackage{pifont}
\usepackage{hyperref}
\hypersetup{colorlinks=true,linktoc=all,linkcolor=blue,breaklinks=true,citecolor=blue,urlcolor=blue}

\usepackage[utf8]{inputenc}
\usepackage[english]{babel}

\definecolor{darkgreen}{RGB}{6, 153, 38}

\addto\captionsenglish{}

\AtBeginDocument{%
    \newwrite\bibnotes
    \def\bibnotesext{Notes.bib}
    \immediate\openout\bibnotes=\jobname\bibnotesext
    \immediate\write\bibnotes{@CONTROL{REVTEX42Control}}
    \immediate\write\bibnotes{@CONTROL{%
    apsrev42Control,author="08",editor="1",pages="1",title="0",year="1"}}
     \if@filesw
     \immediate\write\@auxout{\string\citation{apsrev42Control}}%
    \fi
}%

\newcommand*{\Tr}{\text{Tr}}

\newcommand{\beq}{\begin{equation}}
\newcommand{\eeq}{\end{equation}}

\DeclarePairedDelimiter{\bra}{\langle}{\rvert}
\DeclarePairedDelimiter{\ket}{\lvert}{\rangle}
\DeclarePairedDelimiterX\braket[2]{\langle}{\rangle}{#1\,\delimsize\vert\,\mathopen{}#2}
\DeclarePairedDelimiterX\ketbra[2]{\lvert}{\rvert}{#1\,\delimsize\rangle\mathopen{}\delimsize\langle\,\mathopen{}#2}

\DeclarePairedDelimiterX\Braket[2]{(}{)}{#1\,\delimsize\vert\,\mathopen{}#2}
\DeclarePairedDelimiterX\Ketbra[2]{\lvert}{\rvert}{#1\,\delimsize)\mathopen{}\delimsize(\,\mathopen{}#2}

\newcommand{\kBT}{k_\text{B}T}

\begin{document}

\title{Noninvasive and nonadiabatic quantum Maxwell demon
}
\author{Lucas Trigal}
\affiliation{Instituto de F\'isica Interdisciplinar y Sistemas Complejos IFISC (CSIC-UIB), E-07122 Palma de Mallorca, Spain\looseness=-1}
\affiliation{Departamento de F\'isica Te\'orica de la Materia Condensada, Universidad Aut\'onoma de Madrid, 28049 Madrid, Spain\looseness=-1}
\author{Rafael S\'anchez}
\affiliation{Departamento de F\'isica Te\'orica de la Materia Condensada, Universidad Aut\'onoma de Madrid, 28049 Madrid, Spain\looseness=-1}
\affiliation{Condensed Matter Physics Center (IFIMAC), Universidad Aut\'onoma de Madrid, 28049 Madrid, Spain\looseness=-1}
\affiliation{Instituto Nicol\'as Cabrera (INC), Universidad Aut\'onoma de Madrid, 28049 Madrid, Spain\looseness=-1}
\date{\today}

\begin{abstract}
A quantum mechanical Maxwell demon is proposed in a quantum dot setting. The demon avoids continuous-measurement induced decoherence by exploiting an undetailed charge detector. The control of coherent tunneling via Landau-Zener-St\"uckelberg-Majorana driving allows for efficient feedback operations with no work invested. The local violation of the second law achieves simultaneous power generation and cooling. We discuss the response current fluctuations, and the demon backaction deriving from failures, finding optimal performance in the nonadiabatic regime.  
\end{abstract}

\maketitle


{\it Introduction}---After decades of resting as a conceptual problem understood in terms of information erasure~\cite{landauer:1961,bennett:1982}, the interest on the Maxwell demon ---a mechanism able to decrease the entropy of a system without acting directly on its particles~\cite{leff_maxwellbook}--- has been revived in the last years.
The way has been paved by the development of nanoscale technologies, with their access to microscopic degrees of freedom, bringing proposals and experimental realizations where the action of Maxwell demon devices~\cite{datta:2008,maruyama:2009,whitney_illusory_2023,schaller_how_2024,junior_friendly_2025} can actually be measured. This is so that a rich {\it demonology} has come out in the last years based on different platforms (e.g., quantum dots, ions, qubits), classified depending on whether they need active measurement and feedback on the system~\cite{toyabe_experimental_2010,koski:2014,chida:2017,Ribezzi-Crivellari2019Jul,annbyanderson_maxwell_2020,manzano_thermodynamics_2021,Bhattacharyya2022Dec,aanbyAndersson_maxwell_2024,yan_experimental_2024,aggarwal_rapid_2025} or they work autonomously by means of mutual information flows in nonequilibrium configurations~\cite{whitney:2016,ptaszynski:2018,sanchez:2019,irene,monsel_autonomous_2025}. 
Information heat engines have also been proposed~\cite{hotspots,strasberg:2013,strasberg:2018,erdman_absorption_2018,poulsen_quantum_2022} and realized~\cite{thierschmann:2015,koski:2015} which are based on the exchange of mutual information~\cite{Horowitz2014Jul} but cannot avoid the flow of heat.

\begin{figure}[b]
\includegraphics[width=\linewidth]{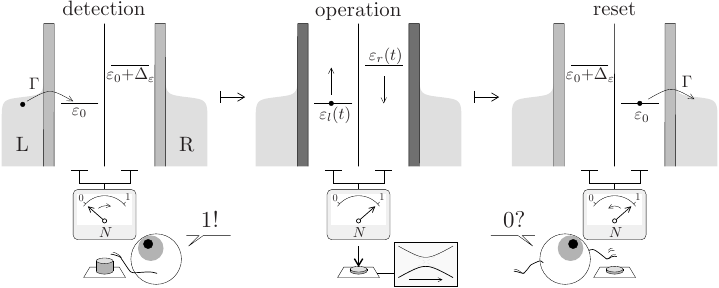}
\caption{\label{fig:scheme}
Three steps of the demon operation inducing transport through a DQD coupled to reservoirs L and R: detection (of whether there is a particle in the DQD), operation (perform a voltage pulse that exchanges the energies of the left and right dot levels), and reset. The demon measures the total charge, $N$, continuously and activates feedback operations when $N$ changes. $\Omega$ is the avoided crossing splitting. 
}
\end{figure}
In the quantum regime, qubit based proposals~\cite{lloyd_quantum_1997,kieu_second_2004,quan_maxwells_2006,quan:2007,elouard_extracting_2017,campisi_feedback_2017,bresque_twoqubit_2021,bhandari_measurement_2023,bresque_runandtumble_2024} and proof of concept experiments~\cite{camati_experimental_2016,cottet_observing_2017,masuyama_information_2018,najerasantos_autonomous_2020,dassonneville_amplifying_2026} use quantum measurements~\cite{andrew_book,landi_current_2024} and unitary gates.  
They involve either local projective measurements (so decoherence and work injection are unavoidable) or that the measurement and the feedback rotation are performed in different bases of the same qubit (so they cannot coexist). 
Here we solve these problems by proposing a Maxwell demon that exploits the quantum nature of the two demonic actions (detection and feedback) combining noninvasive continuous measurements and efficient driven tunneling~\cite{grifoni_driven_1998} in a transport setting. It augments the qubit Hilbert space, ${\bf 1}=\{|l\rangle,|r\rangle\}$, by an empty state $|0\rangle$ that couples it to the work (or heat) extraction reservoirs and allows for continuous detection without perturbing the qubit. 

We exemplify this with a double quantum dot (DQD) charge qubit~\cite{vanderWiel_electron_2002,chatterjee_semiconductor_2021} with interdot coupling $\Omega/2$ mediating the particle flow between two reservoirs (L and R) and coupled to a charge detector (e.g., a quantum point contact~\cite{field_measurements_1993,fujisawa:2006,gustavsson:2006,kung_irreversibility_2012}, a single electron transistor~\cite{haldar_coherence_2024} or a resonator~\cite{petersson_charge_2010,delbecq_coupling_2011,frey_dipole_2012,petersson_circuit_2012,stehlik_fast_2015}), see Fig.~\ref{fig:scheme}. The two states in ${\bf 1}$ are initially energetically split: $\varepsilon_l=\varepsilon_0$ and $\varepsilon_r=\varepsilon_0+\Delta_\varepsilon$, so the occupation of $\ket{l}$ is favored. 
Upon the detection of an electron (most likely coming from L) the demon exchanges the level energies in a time $\tau_d$ (assumed to be much shorter than the characteristic dissipation time scale). 
During the level crossing, the electron tunnels to the other quantum dot with a Landau-Zener-St\"uckelberg-Majorana (LZSM) probability $P_{LZ}$ that is controlled by the driving speed, $\Delta_\varepsilon/\tau_d$~\cite{landau1932,zener1932,stueckelberg1932,Majorana1932}.
By swapping both the dot energies and the charge configuration, the operation conserves energy and thereby enables transport: the electron can tunnel out to $R$. The initial state is then reset.

Provided $\Delta_\varepsilon\gg\Omega$, efficient $|l\rangle\rightarrow|r\rangle$ transfer is possible in the adiabatic ($\tau_d\gg\hbar/\Omega$) regime~\cite{born_beweis_1928} as well as for particular speeds~\cite{shevchenko_lzs_2010,ivakhnenko_nonadiabatic_2023}. 
LZSM tunneling is routinely used in DQDs to enable quantum gates~\cite{hayashi_coherent_2003,petta_coherent_2005,foletti_universal_2009,petta_coherent_2010,dovzhenkononadiabatic_2011,studenikin_quantum_2012,cao_ultrafast_2013,liu_accelerated_2024}, and to probe coherence~\cite{oosterkamp_microwave_1998,petta_manipulation_2004,stehlik_landau_2012,dupontFerrier_coherent_2013,forster_characterization_2014,gonzalez-Zalba_gate_2016,
hecker_coherent_2023},
also in qubits~\cite{nakamura_coherent_1999,oliver_machZehnder_2005,sillanpaa_continuous_2006,wilson_coherence_2007}. 
Note that charge pumps generate transport using either periodic drivings~\cite{pothier_single_1992,buitelaar_adiabatic_2008,roche_two_2013,connolly_gigahertz_2013} without measuring, or conditioned feedback~\cite{kaestner_nonadiabatic_2015}, in both cases with the drive performing work on the system~\cite{juergens_thermoelectric_2013}. To avoid this, the information of when the DQD gets occupied is essential.

The challenge is to measure continuous and noninvasively in a way that the state coherence is not disturbed during the driving operation. 
Protocols conditioned on resolved knowledge of which dot  ($l$ or $r$) is occupied~\cite{annbyanderson_maxwell_2020,aanbyAndersson_maxwell_2024} impede the coherent driving and are affected by heat injection~\cite{ferreira_transport_2024,elouard_revealing_2026,sanchez_virtual_2026}. 
We avoid these drawbacks by imposing that the demon has only a limited access to the state of the system: it detects whether the DQD is empty ($N=0$) or occupied ($N=1$), but not in which quantum dot the electron is. Hence, using the detection operator $\hat{N}_1=\ket{l}\bra{l}{+}\ket{r}\bra{r}$ (the detector is equally coupled to both dots), the coherences between states $|l\rangle$ and $|r\rangle$ are respected during the operation step, even if the detector continues measuring. Note that the total charge $N=\langle \hat{N}\rangle$ plays the role of the memory of the demon in a Szilard engine~\cite{szilard:1929,*szilard_on_1964}. By measuring $N=1$, the demon will hence not be certain that the electron is in the left quantum dot, as desired, but can prepare the system such that this is most likely the case if $\varepsilon_0\ll\varepsilon_0+\Delta_\varepsilon$. 

There is a price for indeterminacy: if the driving goes wrong (with probability $1-P_{LZ}$) the demon performs work and heats the system, which limits its performance far from the adiabatic regime; electrons tunneling from R will be transported by the demon in the wrong direction.

{\it Model}---The DQD is weakly coupled to two fermionic reservoirs at temperature $T$ with tunneling rates $\Gamma\ll\kBT/\hbar$. Strong Coulomb repulsion restricts the system basis to $\{ \ket{0},\ket{l},\ket{r} \}$, with $\hat{d}_\alpha|0\rangle=|\alpha\rangle\in{\bf 1}$. In this regime the electron spin can be ignored, so the Hamiltonian gets: 
\begin{equation}
\hat{H}_{\mathrm{0}}(t)=\sum_{\alpha=l,r}\varepsilon_\alpha(t)\hat{n}_{\alpha}+\frac{\Omega}{2}\left(\hat{d}_{\mathrm{l}}^{\dagger} \hat{d}_{\mathrm{r}}+\hat{d}_{\mathrm{r}}^{\dagger} \hat{d}_{\mathrm{l}}\right).
\end{equation}
The time dependence is due to the demon action, according to the following protocol (see Fig.~\ref{fig:scheme}):  
(i) if $N=0$, $\varepsilon_l=\varepsilon_0$ and $\varepsilon_r=\varepsilon_0+\Delta_\varepsilon$ (detection I stage); (ii) by detecting a  change to $N=1$ at $t=t_k$, the demon swaps the energies of the dots via gate voltages in a time $\tau_d$: $\varepsilon_l(t)=\varepsilon_0+\Delta_\epsilon(t-t_k)/\tau_s$, $\varepsilon_r(t)=\varepsilon_0+\Delta_\epsilon(\tau_dt-t_k)/\tau_d$ (driving stage); (iii) at $t=t_k+\tau_d$, $\varepsilon_l=\varepsilon_0+\Delta_\varepsilon$ and $\varepsilon_r=\varepsilon_0$ is maintained until $N=0$ is detected (detection II stage); (iv) finally a fast drive (compared to $1/\Gamma$) is performed to restore the initial state (reset stage). 
If $\varepsilon_0<\mu_L$ and $\mu_R,\kBT\ll\varepsilon_0+\Delta_\varepsilon$, where $\mu_j$ is the electrochemical potential of reservoir $j=L,R$, tunneling into the DQD from R will be exponentially suppressed by a rate $\propto e^{-(\varepsilon_0+\Delta_\varepsilon-\mu_R)/\kBT}$, so the demon can assume that most electrons come from L. 
To avoid further sources of errors, the demon also decouples the system from the reservoirs during the driving step, which can be done with additional gate voltages. We parametrize this as $\Gamma(t)=\Gamma\chi(t)$, where $\chi=0$ during the driving stage and 1 otherwise. In a successful sequence, the electron tunnels from L to $l$ at $t=t_k$, is transferred to $r$ at $t=t_k+\tau_d$ and then tunnels out to R.
In that case, a particle is transferred across the system, heat $\mu_j-\varepsilon_0$ is extracted from both reservoirs and the demon has performed no work on the system. All the thermodynamic flows can be reversed in the system. The resource is in this case the amount of information generated in the detection apparatus to maintain the continuous measurement \cite{Sagawa2012Nov,Sagawa2013Dec,Horowitz2013Jul,Horowitz2014Jul}.

We assume a fast and errorless detector.
The evolution of the conditioned reduced density matrix of the DQD, $\rho_\gamma$~\cite{gamma}, for a finite time trajectory based on a sequence of measurements records ($\gamma_{(0,\tau)}$) is given by the stochastic master equation~\cite{Manzano2022Jun,landi_current_2024}:
\begin{equation}\label{eq:master_jumps}
    d \rho_\gamma = [H_{0}, \rho_\gamma] \frac{dt}{i} - \chi_\gamma\sum_{i=\pm} \left[  \mathds{M}_{i}(\rho_\gamma) dt - \mathds{J}_i(\rho_\gamma) dN_i \right],
\end{equation}
where $dN_i=\{0,1\}$ is a stochastic increment based on the detection of a transition in or out of the DQD in the interval $[t,t+dt]$.  The first term in Eq.~\eqref{eq:master_jumps} describes the unitary evolution due to $\hat{H}_{0}$. The second dissipative term is introduced by the tunneling events and the detector, given by drift and jump terms:
\begin{align}
\label{eq:M}
\mathds{M}_{i}(\rho_\gamma) & = \sum_{j=L,R} {\frac{1}{2}}\{L^{i\dagger}_{j} L_{j}^i, \rho_\gamma\} - \mathrm{Tr}(L_{j}^{i\dagger} L_{j}^i \rho_\gamma) \rho_\gamma\\
\label{eq:Jump}
\mathds{J}_i(\rho_\gamma) & =\frac{\sum_{j=L,R} L_{j}^i \rho_\gamma L_{j}^{i\dagger}}{\mathrm{Tr}(\sum_{j=L,R} L_{j}^i \rho_\gamma L_{j}^{i\dagger})} - \rho_\gamma,
\end{align}
where $L_{j}^+=\sqrt{\Gamma f_j}\ket{\alpha_j}\bra{0}$ and $L_{j}^-=\sqrt{\Gamma(1{-}f_j)}\ket{0}\bra{\alpha_j}$ are jump operators associated with the tunneling in ($i=+$) and out ($i=-$) of the DQD via reservoir $j$, 
and $f_{j}=[e^{(\varepsilon_{\alpha_j}-\mu_{j}) / \kBT}+1]^{-1}$ is the Fermi-Dirac distribution ($\alpha_L=l$, $\alpha_R=r$), which takes two values: $f_j^0$ and $f_j^\Delta$ depending on whether $\varepsilon_{\alpha_j}=\varepsilon_0$ or $\varepsilon_0+\Delta_\varepsilon$. 
Following Eq.~\eqref{eq:Jump}, an initially empty DQD is updated to a state 
$\rho_{\gamma}'=(f_L \ket{l}\bra{l} +f_R \ket{r}\bra{r} )/(f_L+f_R) $.
In the case $\Delta_{\varepsilon} \gg \mu_R+\kBT$ and $\mu_L -\varepsilon_0\sim\kBT$, the ratio $f_L/f_R\gg1$ guarantees that the protocol will start with a state $\rho_{\gamma}' \approx \ket{L}\bra{L}$, with exponentially suppressed deviations, 
and $1-f_j$ is finite to enable the electron tunneling out to R after the driving. 
Additionally, the condition $\Delta_\varepsilon\ll\varepsilon_0$ suppresses the overlap of the quantum dot states when the system is not being driven, allowing for a local master equation description~\cite{hofer:2017njp}.

\begin{figure}[t]
    \centering
    \includegraphics[width=.9\linewidth]{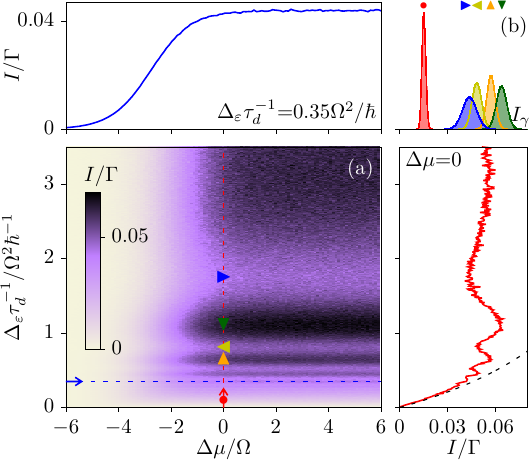}
    \caption{(a) Average particle current through the DQD as a function of the applied electrochemical potential bias and the speed of the energy ramp, with cuts along the marked dotted lines plotted in the lateral panels, with $\Gamma=\Omega/\hbar$, $T_l = T_r = 0.75\Omega/k_{\rm B}$, $\varepsilon_0 = -0.5\Omega$, $\mu_r = 0$, $\Delta = 6\Omega$. The dashed line in the $\Delta\mu=0$ panel plots $[\tau_d+1/\Gamma f^0(1-f^0)]^{-1}$. (b) Histograms of the time-averaged current $I_\gamma\equiv\langle I^\gamma(t)\rangle_t$ of $10^6$ trajectories computed during a time $t_{run}=10^5\hbar/\Omega$ for three different speeds marked by colored symbols in the main panel in (a): $\Delta_\varepsilon/\tau_d=(0.1,0.65,0.8,1.1,1.75)\Omega^2/\hbar$ and $\Delta\mu=0$. 
    We fix the splitting at the avoided crossing, $\Omega\sim\unit[1]{meV}$}.
    \label{fig:current}
\end{figure}
{\it Fluctuating currents}---The action of the demon results in measurable particle and heat currents in the conductor. Each realization of the system fluctuations gives a trajectory of the conditioned $\rho_\gamma(t)$ giving fluctuating particle  and heat currents: 
\begin{align}
\label{eq:partcurr}
 I^\gamma_k(t) \ &= \Gamma^\gamma(t) f_k\rho^\gamma_{00}(t) - \Gamma^\gamma(t)(1-f_k)\rho^\gamma_{\alpha_k\alpha_k}(t)\\
\label{eq:heatcurr}
J^\gamma_k (t) &= (\varepsilon^\gamma_k(t)-\mu_k) I^\gamma_k (t),
\end{align}
with which we can evaluate the rate of entropy change in the reservoirs, $\dot{S}_k^\gamma(t)=-J_k^\gamma(t)/T_k$.

The demon also affects the system energetics: $dE^\gamma_{s}(t) = \operatorname{Tr}[\dot{H}_{0} \rho_\gamma(t)]dt+ \operatorname{Tr}[H_{0}\dot{\rho}_\gamma(t)]dt$, 
where the first term is the work done by the demon during the driving:
\begin{equation}
\label{Wd}
  \dot{W}^\gamma_d (t)  = \frac{\Delta_\varepsilon}{\tau_d}[\rho^\gamma_{ll}(t)-\rho^\gamma_{rr}(t)][1-\chi^\gamma(t)].
\end{equation}
The second term contains the dissipative contributions and the effect of the measurement~\cite{horowitz_quantum_2012,Manzano2022Jun,Elouard2017Mar}, 
appears as fluctuations of the first law \cite{Kerremans2022May} and vanishes upon averaging since $[\hat{H}_{0},\hat{N}]=0$.

The thermodynamic properties are obtained by averaging over all trajectories and over time: $X=\left\langle \langle X^\gamma(t) \rangle_\gamma \right \rangle_t$, with $X\in\{I_j,J_j,\dot{W}_d,\dot{S}_j\}$.  Hence, the first law is
\begin{equation}
\label{eq:1law}
J_L  +  J_R + \dot{W}_d - P  =0,
\end{equation}
with the power, $P=-I\Delta\mu$, being $I\equiv I_L=-I_R$ and $\Delta\mu\equiv\mu_L-\mu_R$. The demon performs useful thermodynamic operations whenever $P>0$ (power generation) or $J_k>0$ (cooling). It is expected from a demon to do this without expending work ($\dot{W}_d=0$) and reducing the system entropy ($\dot{S}_s=\dot{S}_L+\dot{S}_R<0$).

{\it Transport}---The generated particle current, plotted in Fig.~\ref{fig:current}, is robust for a wide range of parameters, even against the bias when $\Delta\mu\leq0$.
In the adiabatic regime, where $\Delta_\varepsilon/\tau_d\to0$, the charge transfer from l to r occurs with a high probability, $P_{LZ}\to1$. Neglecting errors due to tunneling from R, the zero-bias current (so $f^0\equiv f_L^0=f_R^0$) can be estimated by $I(\Delta\mu=0)\approx[\tau_d+1/\Gamma f^0(1-f^0)]^{-1}$~\cite{SM}. Hence the current is suppressed as $1/\tau_d$ for too slow drivings $\tau_d\gg\Gamma^{-1}$, see Fig.~\ref{fig:current}(a). As the system goes faster through the anticrossing, the current exhibits oscillations as a consequence of the coherent LZSM tunneling. The current increases, however it becomes less likely that the transition is done correctly in the nonadiabatic regime and the current saturates. This is reflected in the dispersion of the time averaged trajectories, ${I}_\gamma=\langle I^\gamma(t)\rangle_t$, plotted in Fig.~\ref{fig:current}(b): driving errors done by the demon make the current noisy~\cite{SM}. In the extremely nonadiabatic regime, too many errors suppress the current~\cite{SM}.

When $\Delta\mu<0$, the current flows against the bias and is suppressed as it approaches the condition $\varepsilon_0-\mu_L=\varepsilon_0+\Delta_\varepsilon-\mu_R$, where it is equally probable for electrons to enter the DQD from both reservoirs. As the demon is not able to distinguish the two cases and the driving is symmetric, the current vanishes at this point.
In the opposite regime, $\Delta\mu>0$, the current is constant for fixed $\tau_d$, since $f_L^0\to1$ and $f_R^\Delta\ll1$. There the demon acts as an accelerator as the resulting current is significantly larger than it would be in the absence of the protocol.

\begin{figure}[t]
    \centering
    \includegraphics[width=\linewidth]{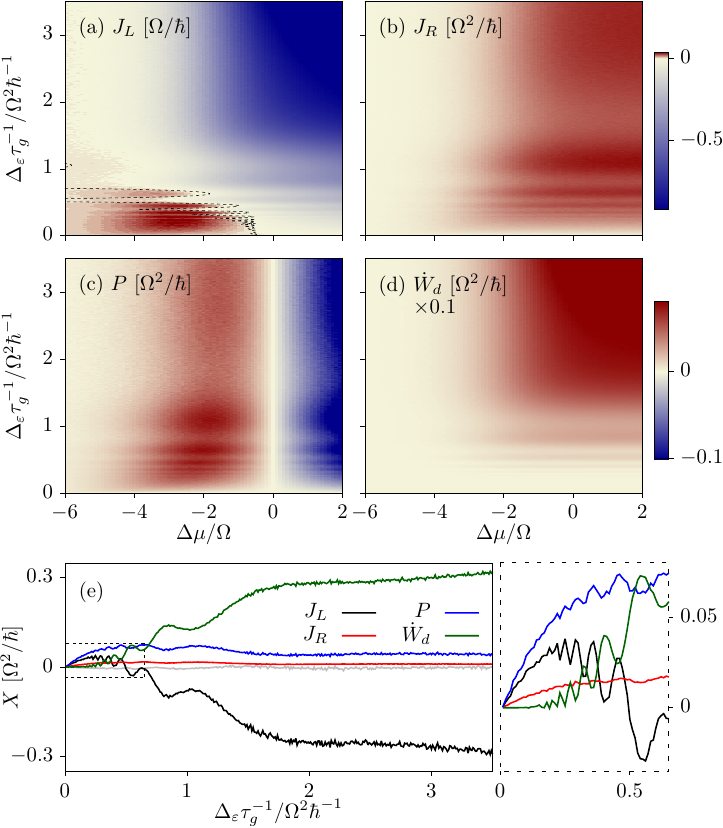}
    \caption{Thermodynamic currents: heat from terminal (a) $L$  and (b) $R$, (c) generated power and (d) work performed by the demon, as functions of the electrochemical potential difference and the driving speed. The dotted line in (a) marks the values of $\mu_L^*$ defined in the text. (e) Cuts of the other panels for fixed $\Delta\mu=-0.22\Omega^2/\hbar$. The grey line represents the left hand side of Eq.~\eqref{eq:1law}. The right panel shows a zoom of the adiabatic region marked by a dotted square. Same parameters of Fig.~\ref{fig:current}.}
    \label{fig:heatcurr}
\end{figure}

The particle current from L to R carries thermal flows which define the regime of operation of the demon, see Fig.~\ref{fig:heatcurr}. 
The discrete quantum dot levels induce cooling whenever electrons are emitted/absorbed by a reservoir over/below its electrochemical potential~\cite{benenti:2017}, which explains the regions with $J_L>0$ (when $\varepsilon_0>\mu_L$ close to the adiabatic regime) and with $J_R>0$ (for a wide range of parameters), cf. Figs.~\ref{fig:heatcurr}(a) and \ref{fig:heatcurr}(b). As anticipated, the system additionally works as a power generator ($P>0$) when $I$ flows against the potential ($\Delta\mu<0$), cf. Fig.~\ref{fig:heatcurr}(c). This means that all thermodynamic flows of the system are reversed. Remarkably in the adiabatic regime, this is done by the demon performing no work, $\dot{W}_d\approx0$, as shown in Figs.~\ref{fig:heatcurr}(d) and \ref{fig:heatcurr}(e): the system has the same energy before and right after the driving stage. 

This ideal performance is damaged with the onset of the nonadiabatic regime as $\Delta_\varepsilon/\tau_d$ increases. Unsuccessful driving events after which the electron does not reach $r$ but instead remains in $l$, induce the electron returning to L at a higher energy $\varepsilon_0 + \Delta_\varepsilon$, therefore not contributing to $I$ but to heat L. Eventually, $J_L<0$ when the demon commits too many errors.
The approximate boundary between these two regimes is given by the set of points for which the energetic contributions due to the demon errors cancel those arising from the correct operation of the protocol out, i.e., when the probability of reaching state $r$ after the driving satisfies:
$\Delta\mu^* = \varepsilon_0 +\Delta_{E}(P_{LZ}-1)/(2P_{LZ}-1)$ as indicated by the dashed line in Fig.~\ref{fig:heatcurr}(a). 
These errors also make $J_R$ and $P$ smaller and force the demon to perform a work $\Delta_\varepsilon$ per unsuccessful drive. As a consequence, the demon cannot avoid to invest $\dot{W}_d>0$ which oscillates in antiphase with the rest of currents (rather than on $P_{LZ}$, it depends on $1-P_{LZ}$~\cite{SM}), see inset in Fig.~\ref{fig:heatcurr}(e). In the strongly nonadiabatic regime, the work performed by the demon dominates all other flows. 
The fulfillment of Eq.~\eqref{eq:1law} is plotted as a grey line in Fig.~\ref{fig:heatcurr}(e). Small deviations from zero reflect the validity of our local description.  

\begin{figure}[t]
    \centering
    \includegraphics[width=.9\linewidth]{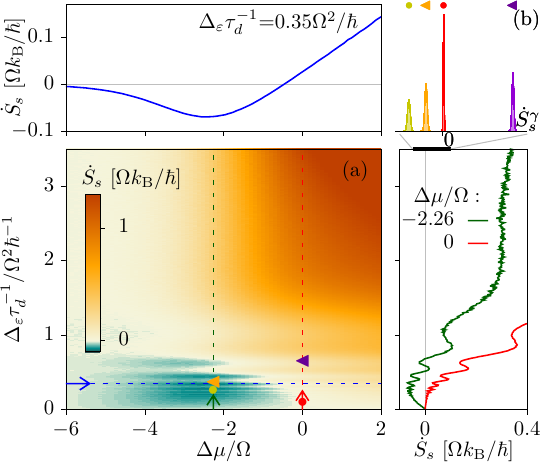}
    \caption{(a) Entropy production in the system reservoirs as a function of the applied electrochemical potential bias and the speed of the energy ramp. Side panels show cuts along the marked dotted lines with corresponding colors. Parameters as in Fig.~\ref{fig:current}. (b) Histograms of the time-averaged entropy production rates, $\dot{S}_s^\gamma=\langle\dot{S}_s^\gamma(t)\rangle_t$ of $10^6$ trajectories computed during a time $t_{run}=10^4\hbar/\Omega$ at the conditions marked by the corresponding symbols in (a): $\Delta_\varepsilon/\tau_d=(0.25,0.36)\Omega^2/\hbar$ and $\Delta\mu=-2.26\Omega$ (yellow and orange), and $\Delta_\varepsilon/\tau_d=(0.1,0.65)\Omega^2/\hbar$ and $\Delta\mu=0$ (red and violet curves).}
    \label{fig:entrop}
\end{figure}

{\it Local violation of the second law}---The entropy production rate in the system due to heat fluxes,
$\dot{S}_s = -J_L/T_L {-} J_R/T_R$, gives a notion of the spontaneity of the flows. If $\dot{S}_s<0$, the heat flows are non-spontaneous, and heat is not flowing in the direction imposed by the formulation of the second law defined by a local observer who only has access to the measurable system currents. This is indeed the problem that motivated Maxwell's thought experiment~\cite{leff_maxwellbook}.
As shown in Fig.~\ref{fig:entrop}(a), this occurs close to the adiabatic regime for $\Delta\mu<0$ (for the chosen parameters), in a region larger than the one delimited by $\Delta\mu^*$ (inside which both $J_L,J_R>0$). Otherwise, the two currents have opposite sign but for some parameters we still find $J_R>-J_L$. Note that the location of the region where $\dot{S}_s<0$ depends on the relative position of $\varepsilon_0$ with respect to the electrochemical potentials $\mu_L$ and $\mu_R$.
The histograms of the time-averaged trajectories show that the demon is so efficient that the whole distribution of values $\dot{S}_s^\gamma\equiv\langle\dot{S}_s^\gamma(t)\rangle_t$ is negative in the region of $\dot{S}_s<0$, see Fig.~\ref{fig:entrop}(b). For $\Delta\mu\ge0$ it is highly improbable to find local violations of the second law, even for very slow drivings.  

For a strict demonic operation, one additionally imposes that the first law is also satisfied locally i.e., $J_L+J_R=P$ and $\dot{W}_d\approx0$. This is the case in the adiabatic regime (low driving speeds), where furthermore the power and heat currents increase with $1/\tau_d$. Their maxima are attained in the weakly nonadiabatic regime. However the demon conditions need to be relaxed there to allow for finite violations of the local first law, as $\dot{W}_d\neq0$. The optimal condition requires a compromise of the amount of extractable power and the invested $\dot{W}_d$.

The full formulation of the second law requires taking into account an informational term associated with the demon, $\dot{S}_d$, due to the measurement and feedback processes. A continuous measurement like the one discussed here requires large information resources and therefore involve a large production of entropy~\cite{Ribezzi-Crivellari2019Jul,Ribezzi-Crivellari2019Aug,Potts2018Nov,Garrahan2023Mar} which will largely overcome the single-electron transport induced reduction of the system entropy. Therefore, the global entropy production, $\dot{S}_s+\dot{S}_d$ will be large and positive. For this reason we do not compute it explicitly in this work. A comprehensive study of the entropic fluxes in the system lies beyond the scope of the present work.

{\it Conclusions}---We have proposed a Maxwell demon that exploits the quantum properties of measurement and feedback in a simple transport setting to ensure the system coherence and energy conservation. Undetailed detection of incoherently coupled subspaces (the total charge) enables a highly probable guess of the system state under appropriate preparation which allows for a half-blind but confident feedback operation. The detector operator commutes with the system Hamiltonian, which averages the backaction out. 
Analyzing the quantum trajectories, we show that the properties of weakly nonadiabatic driving improve the transfer rate and the regularity of the generated current via operations where the demon induces transport performing no work on the system. 
The system entropy is hence reduced (with simultaneous cooling of the two reservoirs and the generation of electrical power) with the first law being respected, the only consumed resource being the continuous measurement.
Compared to classical analogues, the generated current is enhanced while noise is reduced.

Reducing the requirements for detailed knowledge of the system state preserves the system coherence, increases the control of the system energetics (avoiding exchanges inherent to projective measurements on localized states~\cite{liu_maxwell_2026}) and alleviates the need to perform conditioned feedback. 
Fixing the duration of the operation stage 
opens the way for (if not autonomous) automated quantum Maxwell demons by using programmable gates (autonomous bootstrapping)~\cite{zubchenko_autonomous_2025} and detectors~\cite{paurevic_automated_2025}.
Further improvement can be achieved using engineered drivings~\cite{berry_transitionless_2009,brange_adiabatic_2024}, demonstrated in DQDs~\cite{liu_accelerated_2024}, that avoid undesired energy costs~\cite{campbell_tradeoff_2017}.

{\it Acknowledgments}---We thank Bibek Bhandari and Gonzalo Manzano for useful comments on the manuscript. RS is grateful to Tobias Brandes for persistent inspiration. We acknowledge funding from the Spanish Ministerio de Ciencia e Innovaci\'on via grants No. PID2022-142911NB-I00 and No. PID2024-157821NB-I00, and through the ``Mar\'{i}a de Maeztu'' Programme for Units of Excellence in R{\&}D CEX2023-001316-M. L.T. acknowledges financial support from CSIC “JAE PRE” program
(No. JAEPR23015), the CoQuSy project (No. PID2022-140506NB-C21 and C22), and the María de Maeztu Grant (No. CEX2021-001164-M) funded by
MCIU/AEI/10.13039/501100011033 and European Union NextGenerationEU/PRTR.

\bibliography{biblio.bib}

\begin{thebibliography}{119}
\bibitem [{gam()}]{gamma}%
  \BibitemOpen
  \href@noop {} {}\bibinfo {note} {We add a subscript $\gamma$ to
  trajectory-resolved quantities.}\BibitemShut {Stop}%
\end{thebibliography}

\newpage

\setcounter{equation}{0}
\renewcommand{\theequation}{S\arabic{equation}}

\setcounter{figure}{0}
\renewcommand{\thefigure}{S\arabic{figure}}

\onecolumngrid
\vspace{\columnsep}
\section*{SUPPLEMENTARY MATERIAL}
\vspace{\columnsep}
\twocolumngrid



In this supplementary material we give details of the evolution equations in the different stages of the protocol as well as a heuristic model that describes the main features of the numerical simulations in the main text and provides analytical understanding of the involved processes. 

\section*{Protocol steps}

In this appendix, we detail the procedure used to simulate stochastic quantum‐jump trajectories via the Monte Carlo wavefunction method~\cite{landi_current_2024}. We denote by \(\rho_{n-1}\) the state of the double quantum dot (DQD) just after the \((n-1)\)th detection event at time \(t_{n-1}\). Initially, the DQD is empty:
\[
\rho_{0} = \lvert 0\rangle\langle 0\rvert.
\]

\begin{enumerate}
  \item \textbf{No‐jump evolution and sampling the jump time.}\\
    Draw a uniform random number \(r\in[0,1]\). Propagate \(\rho_{n-1}\) under the “no‐jump” superoperator
    \[
      \mathcal{L}_0(\rho)
      = -\,i\bigl[H_{0},\,\rho\bigr]\,dt
        \;-\;\sum_{i=+,-}\!\mathds{M}_{i}(\rho)\,dt,
    \]
    where \(\mathds{M}_{i}(\rho)\) is defined in Eq.~\eqref{eq:M} of the main text. At each time step compute the survival probability
    \[
      P_{\mathrm{surv}}(t)
      = \Tr\bigl\{e^{\mathcal{L}_0\,t}\,\rho_{n-1}\bigr\}.
    \]
    When \(P_{\mathrm{surv}}(\tau_n)=r\), declare \(\tau_n\) the waiting time for the \(n\)th jump, set
    \[
      t_n = t_{n-1} + \tau_n,
      \quad
      \rho_n^* = \frac{e^{\mathcal{L}_0\,\tau_n}\,\rho_{n-1}}
                        {\Tr\{e^{\mathcal{L}_0\,\tau_n}\,\rho_{n-1}\}},
    \]
    and proceed to the jump.
    We do not distinguish which dot becomes occupied—only whether the DQD is empty or not. Thus, immediately after the jump the state is
    \[
      \rho_n
      = \sum_{k=l,r}
        \frac{L^{\mathrm{in}}_k\,\rho_n^*\,\bigl(L^{\mathrm{in}}_k\bigr)^\dagger}
             {\Tr\{L^{\mathrm{in}}_k\,\rho_n^*\,\bigl(L^{\mathrm{in}}_k\bigr)^\dagger\}}
      = \frac{f_l\,\lvert l\rangle\langle l\rvert \;+\; f_r\,\lvert r\rangle\langle r\rvert}
             {f_l + f_r},
    \]
    where $L^{\mathrm{in}}_k
      = \sqrt{\Gamma\,f_k(\varepsilon_k)}\;\lvert k\rangle\langle 0\rvert$.

  \item \textbf{Decoupling and driven evolution.}\\
    Immediately after detection, switch off tunneling (\(\Gamma\to0\)) and sweep the dot energies linearly through the avoided crossing. The state \(\rho_n\) then evolves under
    \[
      \dot\rho = \mathcal{L}_d(\rho) = -\,i\bigl[H_{0}(t),\,\rho\bigr].
    \]
    Integrating for a duration \(\tau_d\) gives
    \[
      \rho_n(t_n+\tau_d)
      = \frac{e^{\mathcal{L}_d\,\tau_d}\,\rho_n(t_n)}
               {\Tr\{e^{\mathcal{L}_d\,\tau_d}\,\rho_n(t_n)\}}.
    \]

  \item \textbf{Re‐coupling and the next jump.}\\
    Re‐enable tunneling to the reservoirs and repeat steps 1–3, now starting from the occupied state \(\rho_n(t_n+\tau_d)\). Draw a new random number \(r'\), propagate under \(\mathcal{L}_0\) until
    \(\Tr\{e^{\mathcal{L}_0\,\tau_{n+1}}\,\rho_n(t_n+\tau_d)\}=r'\), record \(\tau_{n+1}\), and reset the DQD to the empty state \(\ket{0}\langle 0\rvert\).
\end{enumerate}

\section*{Heuristic model}

We can write a simple model based on a rate equation for sequential transitions between an enlarged Hilbert space. It considers the empty state $\ket{0}$, and $\ket{l}$, $\ket{r}$ for states before the driving, and $\ket{l^*}$, $\ket{r^*}$ for states after the driving. The successful transitions $\ket{l}\to\ket{r^*}$ and $\ket{r}\to\ket{l^*}$ have a rate $\gamma_{LZ}=P_{LZ}/\tau_d$, and the errors $\ket{l}\to\ket{l^*}$ and $\ket{r}\to\ket{r^*}$ a rate $\bar\gamma_{LZ}=(1-P_{LZ})/\tau_d$. Assuming that the detection of tunneling through the contacts to the reservoirs is instantaneous, the transitions $\ket{L}\to\ket{0}$ and $\ket{R}\to\ket{0}$ are avoided, hence explicitly breaking detailed balance. We furthermore assume that the splitting is large enough for the interdot overlap to be negligible during the detection stages, $\Delta_\varepsilon\gg\Omega$~\cite{sprekeler_coulomb_2004}. Then a local description in terms of localized states is convenient~\cite{hofer:2017njp}. 

\begin{figure}[t]
    \centering
    \includegraphics[width=.9\linewidth]{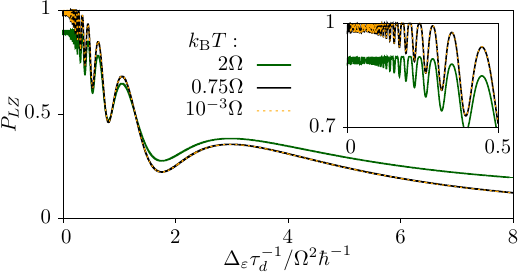}
    \caption{Occupation probability of the right dot after going through the avoided crossing at a speed $\Delta_\varepsilon/\tau_d$, for $\varepsilon_0=-0.5\Omega$ and $\Delta_\varepsilon=6\Omega$, for three different temperatures of the reservoirs, $\kBT=(10^{-3},0.75,2)\Omega$, and $\Delta\mu=0$. The lower temperature curves are almost identical. The inset zooms in the adiabatic regime. }
    \label{fig:plz}
\end{figure}

The transfer probability, $P_{LZ}$, is obtained by propagating the initial state $\hat\rho(t_0)=(f_L^0,f_R^\Delta)\mathbb{1}_2/(f_L^0+f_R^\Delta)$ during the drive stage via the von Neumann equation $\dot{\hat{\rho}}(t)=[\hat{H}_{0}(t),\hat\rho(t)]/i\hbar$, such that $P_{LZ}\equiv\rho_{rr}(t_0+\tau_d)$. The Fermi functions $f_j^0=1/\{1+\exp[(\varepsilon_0-\mu_j)/\kBT]\}$ and $f_j^\Delta=1/\{1+\exp[(\varepsilon_0+\Delta_\varepsilon-\mu_j)/\kBT]\}$ give the occupation of the reservoirs at the tunneling energies. The result is plotted in Fig.~\ref{fig:plz} for two different temperatures of the system at $\Delta\mu=0$. As temperature increases, the initial occupation of $\ket{r}$ reduces $P_{LZ}$ and hence the confidence of the demon in the driving.

With this and the Fermi golden rule transition rates $\Gamma_{L0}^+=\Gamma_Lf_L^0$ and $\Gamma_{R\Delta}^+=\Gamma_Rf_R^\Delta$ for tunneling in (via transitions $\ket{0}\to\{\ket{l},\ket{r}\}$, respectively), and $\Gamma_{L\Delta}^-=\Gamma_L(1-f_L^\Delta)$ and $\Gamma_{R0}^-=\Gamma_R(1-f_R^0)$ for tunneling out (via $\{\ket{l^*},\ket{r^*}\}\to\ket{0}$), the master equation for the occupations $p_j=\rho_{jj}$ reads:
\begin{align}
\label{eq:heurmastereq}
\dot{p}_{0}&=\Gamma_{R0}^-p_{r^*}+\Gamma_{L\Delta}^-p_{l^*}-(\Gamma_{l0}^++\Gamma_{R\Delta}^+)p_{0}\nonumber\\
\dot{p}_{l}&=\Gamma_{L0}^+p_{0}-\tau_d^{-1}p_{l}\nonumber\\
\dot{p}_{r}&=\Gamma_{R\Delta}^+p_{0}-\tau_d^{-1}p_{r}\\
\dot{p}_{l^*}&=\bar\gamma_{LZ}p_{l}+\gamma_{LZ}p_{r}-\Gamma_{L\Delta}^-p_{l^*}\nonumber\\
\dot{p}_{r^*}&=\gamma_{LZ}p_{l}+\bar\gamma_{LZ}p_{r}-\Gamma_{R0}^-p_{r^*},\nonumber
\end{align}
where $\Gamma_L$ and $\Gamma_R$ are the transparencies of the barriers connecting the DQD to reservoirs L and R.

\begin{figure}[t]
    \centering
    \includegraphics[width=.9\linewidth]{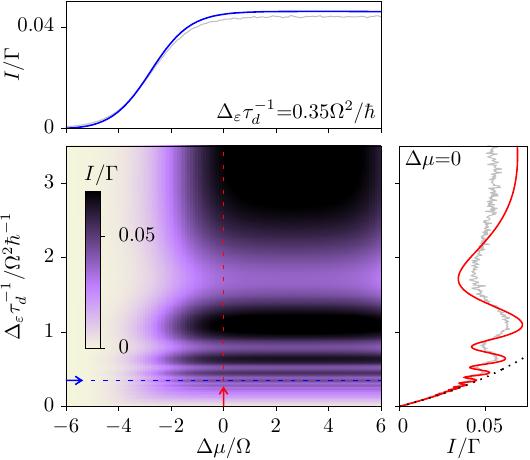}
    \caption{Average particle current through the DQD  computed with the heuristic model as a function of the applied electrochemical potential bias and the speed of the energy ramp, with cuts along the marked dotted lines plotted in the lateral panels. $\Gamma = \Omega/\hbar$, $T_l = T_r = 0.75\Omega/k_{\rm B}$, $\varepsilon_0 = -0.5\Omega$, $\mu_r = 0$, $\Delta = 6\Omega$. In the lateral panels, the model is compared to the numerical results in Fig.~\ref{fig:current} of the main text (grey lines). The black dotted line in the rightmost panel plots Eq.~\eqref{eq:adiabcurr}. }
    \label{fig:currheu}
\end{figure}

In the stationary regime, $\dot{p}_j=0$, the particle current, $I=\Gamma_{R0}^-p_{r^*}-\Gamma_{R\Delta}^+p_0$, becomes:
\begin{equation}
\label{eq:heurcurr}
I=\frac{P_{LZ}(\Gamma_{L0}^+-\Gamma_{R\Delta}^+)\Gamma_{L\Delta}^-\Gamma_{R0}^-}{\tau_d\Lambda_3},
\end{equation}
where $\Lambda_3\equiv(\tau_d^{-1}+\Gamma_{R\Delta}^++\Gamma_{L0}^+)\Gamma_{L\Delta}^-\Gamma_{R0}^-+\tau_d^{-1}[\Gamma_{R0}^-\Gamma_{L0}^++\Gamma_{L\Delta}^-\Gamma_{R\Delta}^++P_{LZ}(\Gamma_{L0}^+-\Gamma_{R\Delta}^+)(\Gamma_{L\Delta}^--\Gamma_{R0}^-)]$.
The second term within parenthesis in Eq.~\eqref{eq:heurcurr} accounts for unwanted transitions that are undetectable to the demon. However, in the appropriate configuration with $\varepsilon_0+\Delta_\varepsilon\gg\mu_R,\kBT$, we have $\Gamma_{L0}^+\gg\Gamma_{R\Delta}^+$. Additionally, in the adiabatic regime, $\gamma_{LZ}\to\tau_d^{-1}$ and $\bar\gamma_{LZ}\to0$, resulting in an error-free current:
\beq
\label{eq:adiabcurr}
I_{\rm e-f}=\left(\tau_s+\frac{1}{\Gamma_{L0}^+}+\frac{1}{\Gamma_{R0}^-}\right)^{-1}.
\eeq
The current in Eq.~\eqref{eq:heurcurr} is plotted in Fig.~\ref{fig:currheu} and compared with the numerical results shown in Fig.~\ref{fig:current} of the main text for the same parameters, with very good agreement. The low speed linear dependence reproduces the result of Eq.~\eqref{eq:adiabcurr}, see black dotted line in the rightmost panel of Fig.~\ref{fig:currheu}. 

\subsection*{Thermal currents}

\begin{figure}[t]
    \centering
    \includegraphics[width=\linewidth]{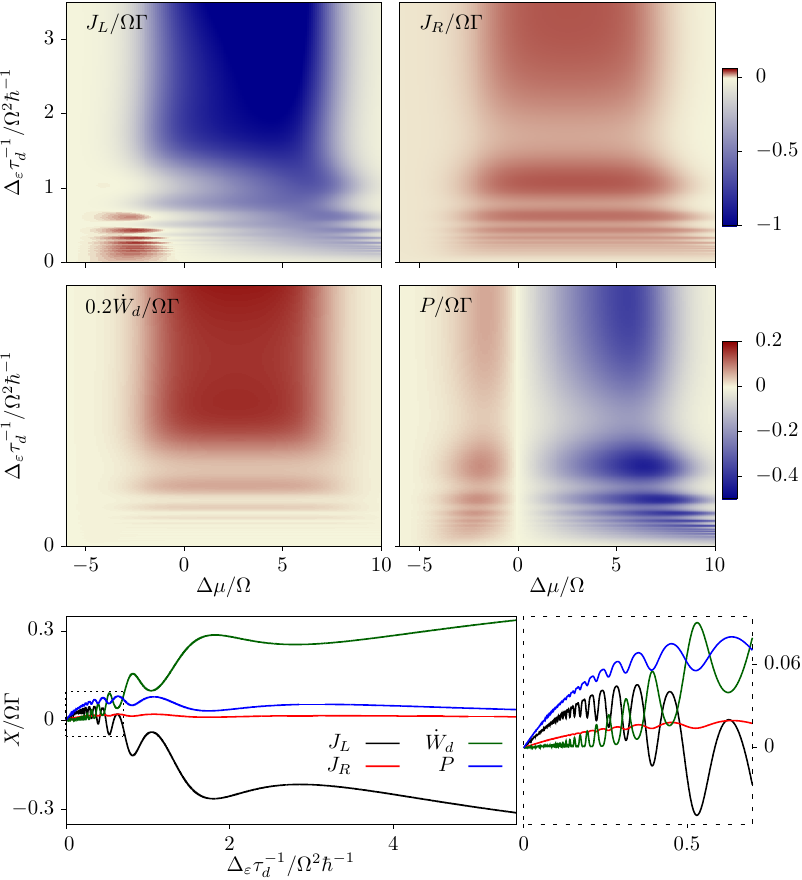}
    \caption{Thermal currents $J_L$, $J_R$, $\dot{W}_d$ and $P$ as functions of the electrochemical potential bias and the driving speed, computed with the heuristic model for the same parameters used in Fig.~\ref{fig:currheu}. The lower panels plot cuts for fixed $\Delta\mu=-2.25\Omega$, the rightmost one zooming in the dotted square region.}
    \label{fig:heatheu}
\end{figure}

We can also calculate the heat currents, $J_L=(\varepsilon_0-\mu_L)\Gamma_{L0}^+p_0-(\varepsilon_0+\Delta_\varepsilon-\mu_L)\Gamma_{L\Delta}^-p_{l^*}$ and $J_R=(\varepsilon_0+\Delta_\varepsilon-\mu_R)\Gamma_{R\Delta}^+p_0-(\varepsilon_0-\mu_R)\Gamma_{R0}^-p_{r^*}$, the generated power, $P=(\mu_R-\mu_L)I$, and the work performed by the demon due to the unsuccessful driving events, $\dot{W}_d=\Delta_\varepsilon\bar\gamma_{LZ}(p_l-p_r)$. The resulting expressions
\begin{align}
\label{Seq:jlheur}
J_L&=\frac{(\varepsilon_0{-}\mu_L)\Gamma_{L0}^+-(\varepsilon_1{-}\mu_L)\Gamma_{R\Delta}^+-\eta_{LZ}^{}\Delta_\varepsilon\Gamma_{L0}^+}{\Gamma_{L0}^+-\Gamma_{R\Delta}^+}I\\
\label{Seq:jrheur}
J_R&=\frac{(\varepsilon_1{-}\mu_R)\Gamma_{R\Delta}^+-(\varepsilon_0-\mu_L)\Gamma_{L0}^++\eta_{LZ}^{}\Delta_\varepsilon\Gamma_{R\Delta}^+}{\Gamma_{L0}^+-\Gamma_{R\Delta}^+}I\\
\label{Seq:wdheur}
\dot{W}_d&=\eta_{LZ}^{}\Delta_\varepsilon I
\end{align}
explicitly verify the first law $J_L+J_R+\dot{W}_d=P$,
where $\varepsilon_1\equiv\varepsilon_0+\Delta_\varepsilon$ and $\eta_{LZ}^{}=(1-P_{LZ})/P_{LZ}$ is the driving error rate.

\begin{figure}[b]
    \centering
    \includegraphics[width=\linewidth]{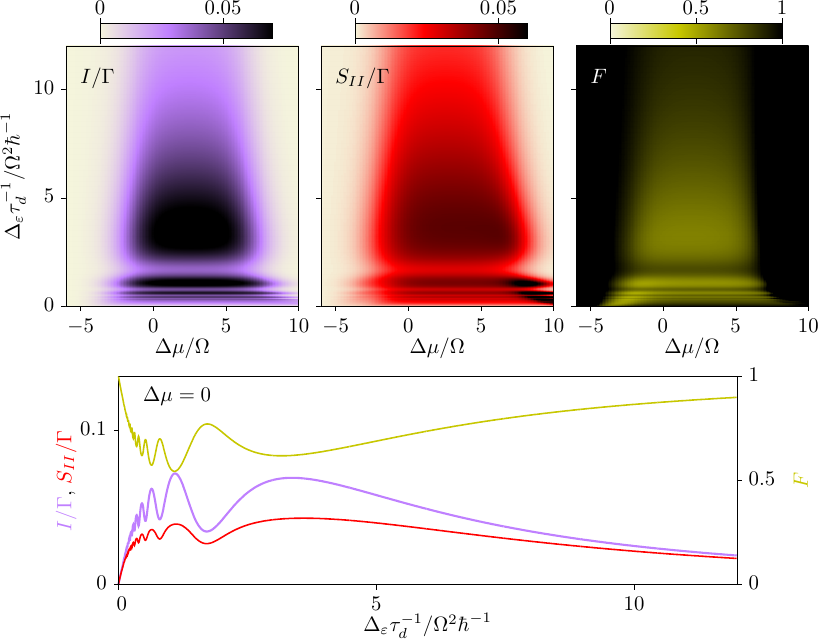}
    \caption{Particle current, noise and Fano factor, $F=S_{II}/I$, computed with the heuristic model for the same parameters as in Fig.~\ref{fig:currheu}. The lower panel shows zero-bias cuts of the above quantities. }
    \label{fig:noiseheur}
\end{figure}

In Fig.~\ref{fig:heatheu} we plot the thermal currents. The model reproduces all the features described in the main text, including the cooling in L and R, power generation for $\Delta\mu<0$ and the suppression of the driving work in the adiabatic regime. For large and positive $\Delta\mu$, the particle current is suppressed (see Fig.~\ref{fig:noiseheur}): if $\mu_L>\varepsilon_0+\Delta_\varepsilon$, the electron remaining in dot $l$ after a failed driving stage has an exponentially suppressed rate to tunnel out to L and therefore blocks the current (in this simplified model we are neglecting interdot tunneling after the drive).  

\subsection*{Efficiency}

\begin{figure}[t]
    \centering
    \includegraphics[width=.9\linewidth]{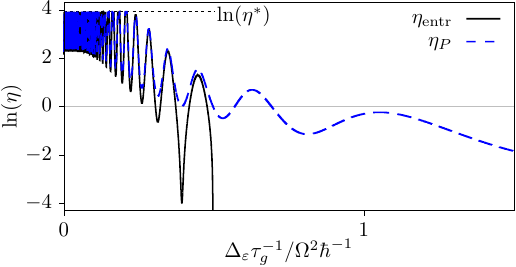}
    \caption{Efficiency of the entropy reduction, $\eta_{\rm entrop}$, and power generation, $\eta_P$ processes as functions of the driving speed. The upper bound $\eta^*$ is given in the text.}
    \label{fig:effheu}
\end{figure}

With Eqs.~\eqref{Seq:jlheur}--\eqref{Seq:wdheur} we define different efficiencies, depending on whether the focus being put on violations of the second law:
\begin{equation}
\eta_{\rm entr}\equiv\frac{-\dot{S}_sT}{\dot{W}_d}
\end{equation}
or on power generation:
\begin{equation}
\eta_{P}\equiv\frac{P}{\dot{W}_d}.
\end{equation}
In the isothermal case ($T_L=T_R=T$), we get from energy conservation:
\beq
\eta_{\rm entr}=\eta_{P}-1,
\eeq
with $\eta_P=(\mu_R-\mu_L)/\eta_{LZ}\Delta_\varepsilon$. In this sense, $\eta_{LZ}\Delta_\varepsilon$ defines the {\it probabilistic}
resource expenditure of the demon. 
The demonic operations will correspond to having $\eta_{\rm entr}>1$ or $\eta_P>1$. In the adiabatic regime, $\eta_{LZ}\to0$ makes the efficiencies huge and bounded by the system thermodynamic properties via the initial condition of the driving. 
Estimating that $P_{LZ}^{\rm ad}\approx f_L^0/(f_L^0+f_R^\Delta)$, we find the upper bound $\eta_P\leq\eta^*=(\mu_R-\mu_L)f_L^0/\Delta_\varepsilon f_R^\Delta$, shown in Fig.~\ref{fig:effheu}. 
Errors in the driving reduce the efficiencies further. When they are below 1, the the demon operates correspondingly as a measurement-enabled heat engine or power transducer. For large driving speeds, entropy production becomes positive and the entropic demon stops working, while the power generation becomes increasingly inefficient.

\subsection*{Noise}
Using the heuristic master equation \eqref{eq:heurmastereq}, we can also calculate the zero frequency current noise, $S_{II}$ using full counting statistics techniques~\cite{sanchez_resonance_2007}. The signal to noise ratio is given by the Fano factor, $F=S_{II}/I$. Superpoissonian regions (due to thermal fluctuations when transport is suppressed) are plotted in black. The result, plotted in Fig.~\ref{fig:noiseheur}, confirms that the zero-bias current is not only higher but also more regular in the intermediate regime. It becomes Poissonian in the adiabatic and highly nonadiabatic regimes, but for different reasons: in the adiabatic regime, the charge fluctuations are unidirectional and dominated by the coupling to the reservoirs, with transport being suppressed by the very long driving times; in the highly nonadiabatic regime, the driving becomes noisy and only few particles are successfully transferred to the right reservoir.

\end{document}